\documentclass[fleqn,12pt]{article}
\usepackage{amsmath,amsfonts,epsfig}

\def\be{\begin{equation}}       \def\eq{\begin{equation}}
\def\ee{\end{equation}}         \def\eqe{\end{equation}}

\def\bea{\begin{eqnarray}}      \def\eqa{\begin{eqnarray}}
\def\ena{\end{eqnarray}}        \def\eea{\end{eqnarray}}
                                \def\eqae{\end{eqnarray}}

\def\a{\alpha}
\def\b{\beta}
\def\c{\gamma}
\def\d{\delta}

\def\s{\sigma}                                   

\newcommand{\Tr}{\mbox{Tr}}

\newcommand{\adot}{{\dot{\alpha}}}

\newcommand{\bdot}{{\dot{\beta}}}

\begin{document}

\begin{titlepage}

\begin{flushright}
IHES/P/05/48
\end{flushright}

\vspace{1cm}
\begin{center}

{\bf A Chiral Perturbation Expansion for Gravity }

\vspace{1cm}

M.\ Abou-Zeid\footnote{abouzeid@tena4.vub.ac.be} and C.\ M.\
Hull\footnote{c.hull@imperial.ac.uk}

\vspace{.5cm} $^1${\small \em Theoretische Natuurkunde, Vrije
Universiteit Brussel \& The International Solvay
Institutes,\\Pleinlaan 2, 1050 Brussels, Belgium}

\vspace{.5cm} $^1${\small \em Institut des Hautes Etudes Scientifiques,\\35 route de Chartres, Le Bois-Marie, 91440 Bures-sur-Yvette, France}

\vspace{.5cm} $^2${\small \em Theoretical Physics Group, The Blackett
Laboratory, Imperial College London,\\ Prince Consort Road, London
SW7 2BW, United Kingdom}

\vspace{.5cm} {\small \em $^2$The Institute for Mathematical
Sciences, Imperial College London,\\48 Princes Gardens, London SW7 2AZ, United Kingdom} \\

\vspace{1cm}

\begin{abstract}
\noindent A formulation of
 Einstein gravity, analogous to that for gauge
theory arising from the Chalmers-Siegel action,
leads to a perturbation theory  about an asymmetric weak coupling limit that treats positive and negative helicities differently.
 We find power counting rules for amplitudes that suggest the theory
 could find a natural interpretation in terms of a twistor-string
theory for gravity with amplitudes supported on holomorphic curves in twistor space.
\end{abstract}

\end{center}
\vspace{1cm}

\flushleft{November 2005}

\end{titlepage}

Tree-level MHV amplitudes for (super)Yang-Mills
theory~\cite{PT,BG} have an elegant formulation in twistor
space~\cite{Nair1}, and Witten considered the extension of  this
to general amplitudes in~\cite{Witten2003}, where it was
conjectured that   amplitudes are non-zero only if all the
external particles in a scattering process are represented by
points in twistor space that lie on an algebraic curve of degree
$d$ given by
\begin{equation}
d = q-1 +l ,\label{Wittform}
\end{equation}
where $q$ is the number of negative helicity particles and $l$ is
the number of loops. This can be understood as resulting from an
underlying twistor string theory~\cite{Witten2003,NB,BM} and
twistor string theory also leads to conformal (super)gravity,
where similar results apply~\cite{BWsc}. There has since been
great progress in understanding general super-Yang-Mills amplitudes in twistor
space; see e.g.~\cite{CachSv} and references therein.

The twistor strings of~\cite{Witten2003,NB,BM} have the problem
that conformal supergravity is inextricably mixed in with the gauge
theory, so that   conformal
supergravity modes propagate on internal lines in gauge theory loop amplitudes
and there appears to be no decoupling limit giving pure super-Yang-Mills amplitudes. A twistor string
that gave Einstein supergravity coupled to super Yang-Mills would
be much more useful, and might have a limit in which the gravity
could be decoupled.

It is known that MHV amplitudes for Einstein (super)
gravity~\cite{BGK} also have an elegant formulation in twistor
space~\cite{Witten2003,Gio,Nair2,Bjeretal}. Our purpose here is to
seek further evidence that (super)gravity amplitudes could arise
from a twistor string theory. An interesting way of
understanding~(\ref{Wittform})  in gauge theory~\cite{Witten2003}
is that it follows naturally  from the perturbation theory of the
Chalmers-Siegel  chiral formulation  of Yang-Mills
theory~\cite{CS}, in which positive and negative helicities are
treated very differently. Moreover, the Chalmers-Siegel
formulation is precisely the form of the gauge theory that arises
from the perturbative twistor string of ref.~\cite{Witten2003}. We
will investigate here chiral formulations of gravity, and show
that perturbation theory about them  again leads to the
relation~(\ref{Wittform}), suggesting that such a chiral
formulation of gravity might arise from a twistor string.

The   action for Yang-Mills can be written as
\begin{equation}\label{SiegelYM2}
\int d^4 x   \Tr \left(G^{\mu\nu} F_{\mu\nu} - \frac{\epsilon}{2}
G^{\mu\nu}G_{\mu\nu}  \right) ,
\end{equation}
where $F=dA+A\wedge A$ is the Yang-Mills field strength and
$G=*G$ is a self-dual auxiliary 2-form taking values in
the   gauge algebra.
Eliminating the auxiliary field  $G_{\mu\nu}$ from~(\ref{SiegelYM2}) gives the
 Yang-Mills action
\begin{equation}\label{SiegelYM3}
\frac{1}{2\epsilon}\int d^4 x   \Tr \left( F^{(+)}_{\mu\nu}
F^{(+)\mu\nu}   \right) ,
\end{equation}
where $F^{(\pm)}$ are the self-dual and
anti-self dual parts of the field strength:
 $F^{(\pm)}= \frac{1}{2} (F \pm *F)$ in signatures $(2,2)$ and $(4,0)$, or
 $F^{(\pm)}= \frac{1}{2} (F \pm i*F)$ in Lorentzian signature. Here we will present formulae for signatures
 $(2,2)$ and $(4,0)$ so that all fields are real; the generalisation to signature $(3,1)$ is straightforward, but involves complex actions.
 The action~(\ref{SiegelYM2}) is
 \begin{equation}\label{YMaction}
 \frac{1}{4g_{YM}^2}\int \,  \Tr \left( F\wedge * F \right)   +\frac{1}{4g_{YM}^2}\int \,  \Tr \left( F\wedge  F
 \right) ,
 \ee
where  $g_{YM}^2=\epsilon/2$.
The first term is the usual Yang-Mills action, and the second is a topological term proportional
to the 2nd Chern number that does not affect the equations of motion or perturbation theory.

The weak coupling limit given by setting $\epsilon =0$
in~(\ref{SiegelYM2}) gives
 Siegel's chiral theory~\cite{Siegel1}
(see also~\cite{Parkes})  with  action
\begin{equation}\label{SiegelYM1}
\int   \Tr \left( G \wedge F \right) = \int d^4 x \Tr
\left(G \wedge F^{(+)} \right) .
\end{equation}
The
field $G$ is a Lagrange multiplier field whose variation gives the
 constraint
\begin{equation}\label{condFplusiszero}
  F^{(+)} (A) = 0 ,
\end{equation}
  implying $F=F^{(-)}$, so that the field strength is
anti-self-dual.
The field equation obtained by varying $A$ is
\begin{equation}
D^\mu  G_{\mu\nu} = 0  ,\label{Gfeq2}
\end{equation}
where $D_\mu = \partial _\mu-A_\mu$ is the gauge covariant derivative.
Eq.~(\ref{Gfeq2})  is of the same form as the Yang-Mills equation
$D^\mu F_{\mu\nu} = 0$. The theory describes
 a helicity $+1$ particle represented by the Yang-Mills  field $A$
with field strength satisfying
(\ref{condFplusiszero}) (so that
$F=F^{(-)}$ satisfies $D^\mu  F^{(-)}_{\mu\nu}
= 0$) and a helicity $-1$ particle represented by the independent
field $G^{(+)}$ satisfying $D^\mu  G^{(+)}_{\mu\nu} = 0$. The
linearized spectrum is  the same for~(\ref{SiegelYM1})
and~(\ref{YMaction}), but the interactions are different: the
action~(\ref{SiegelYM1}) has an $AAG$ term, describing a vertex of
three fields with helicities $++-$, but, in contrast to Yang-Mills
theory, it has no $--+$ vertex.

 The   theory with action~(\ref{SiegelYM1})
has the same spectrum as   Yang-Mills theory, viz.\
particles of helicities +1 (represented by $A$) and $-1$ (represented by
$G$), but differs in the interactions, and it is a non-trivial weak coupling
limit of the standard  theory written in the
form~(\ref{SiegelYM2}). Perturbation theory in $\epsilon$ based on
the action~(\ref{SiegelYM2}) is an expansion about Siegel's theory~(\ref{SiegelYM1}) and treats positive and negative
helicity gluons rather differently.

It is useful to attribute to
the independent fields $A$ and $G$ the weights $w[A]=0 $ and
$w[G]=-1$ under a $U(1) $ transformation, related to the
  \lq anomalous' $U(1)$ R-symmetry $S$
 in the $N=4$ supersymmetric extension of
action~(\ref{YMaction})~\cite{Witten2003}, with $S=4w$. The
Yang-Mills action~(\ref{YMaction}) then has weight $w=0$ while the
Siegel action~(\ref{SiegelYM1}) has weight $w=-1$ and the second
term in~(\ref{SiegelYM2}) has weight $w=-2$. In~\cite{Witten2003},
the $w=-1$ term~(\ref{SiegelYM1}) was interpreted as the
 transform  to space-time of holomorphic Chern-Simons theory on twistor
space, while the $w=-2$ interaction term was related to
nonperturbative D-instanton contributions in twistor-string
theory.

Consider the perturbation theory in $\epsilon$ for the
action~(\ref{SiegelYM2}), which was analysed in~\cite{Witten2003}.
If one attributes weights $w=1$ to $\epsilon$ and $w=-1$ to the
Planck constant $\hbar$, then the action rescaled by $1/ \hbar$
has weight $w=0$. The generating functional of
  scattering matrix elements at $l$-loops must be a sum of terms of the form
  \be
\hbar^{l-1} f(A)  \epsilon^d G^q \ee for some function $f$, and
for this to have total weight zero it is necessary that the
relation~(\ref{Wittform}) holds for each term in the effective
action~\cite{Witten2003}. The power $q$ of $G$ is the number of
negative helicity gluons in an $l$-loop scattering process, while the power $d$ of
$\epsilon$ is the instanton number, given by the degree of the
holomorphic curve in twistor space.

We now turn to gravity, formulated in terms of  a vierbein
$e^{a}{}_\mu$ and   spin-connection $\omega_{\mu}{}^{bc}$, with corresponding one-forms $e^a,\omega^{bc}$.
The torsion and   curvature  2-forms
are given by
\begin{equation}\label{Texpl}
  T^a =   d e^a +
  \omega^a{}_b \wedge
  e^b
   \end{equation}
and
\begin{equation}\label{R2form}
R^a{}_b (\omega )=d \omega^a{}_b + \omega^a{}_c \wedge w^c{}_b .
\end{equation}
In a second order formalism, one imposes the constraint
\begin{equation}\label{Tiszero}
  T^a = 0 ,
\end{equation}
 which
determines the spin-connection in terms of the vierbein:
\begin{equation}\label{omandc}
  \omega_{\mu ab} =  \Omega_{\mu ab} (e) .
\end{equation}
Here $\Omega_{\mu ab} (e)$ is the usual expression for the Lorentz
connection in terms of the vierbein,
\begin{equation}
 \Omega_{\mu}{}^{ab} (e) \equiv  e^{\nu a} \partial_{[\mu}
 e_{\nu ]}{}^b -e^{\nu b} \partial_{[\mu}
 e_{\nu ]}{}^a -e^{\rho a }e^{\sigma b} \partial_{[\rho}
 e_{\sigma ] c}e_{\mu}{}^c .
 \label{usom}
\end{equation}
The Einstein-Hilbert action is
\begin{equation}
\frac{1}{4\kappa^2} \int    e^a \wedge e^b \wedge R^{cd} (\omega
) \varepsilon_{abcd} . \label{EH1order}
\end{equation}
The same action can   be used in the first order formalism, in which the torsion is unconstrained and the
  vierbein $e_{\mu}{}^a$ and the connection $\omega_{\mu}{}^{ab}$ are treated as independent
variables.
The field equation  obtained by varying  $\omega$
is~(\ref{Tiszero}),  which implies that the Lorentz connection is
the Levi-Civita connection~(\ref{omandc}). The vielbein field
equation then gives the Einstein equation.

In Euclidean signature $(4,0)$, the spin group factorises as $Spin(4)=SU(2)\times SU(2)$
while in split signature it factorises as $Spin(2,2)=SU(1,1)\times SU(1,1)$. The spin-connection
decomposes into the self-dual piece $\omega^{(+)ab}$ and
the anti-self-dual piece $\omega^{(-)ab}$,
\begin{equation}\label{omegapm}
  \omega^{(\pm )}_{bc }\equiv \frac{1}{2} \left( \omega_{bc} \pm \frac{1}{2}
  \varepsilon_{bc}{}^{de} \omega_{de} \right) ,
\end{equation} which are the independent gauge fields for the two factors of the spin group.

The curvature 2-form can also be split into  self-dual and anti-self-dual pieces
\begin{equation}\label{Rpm}
  R^{(\pm )}_{bc }\equiv \frac{1}{2} \left( R_{bc} \pm \frac{1}{2}
  \varepsilon_{bc}{}^{de} R_{de} \right) ,
\end{equation}
and it is easily seen that $R^{(+)ab}$ depends only on $\omega^{(+)}$
while $R^{(-)ab}$ depends only on $\omega^{(-)}$, with
 \begin{equation} \label{Rsd}
R^{(\pm )a}{}_b (\omega )=d \omega^{(\pm )a}{}_b +  \omega^{(\pm
)a}{}_c \wedge \omega^{(\pm )c}{}_b .
\end{equation}
In 2-component spinor notation, where $\a,\b$ transform under the first $SU(2)$ or $SU(1,1)$ factor and
$\adot, \bdot$ transform under the second,
$\omega^{(+)ab}$ becomes $\omega^{\alpha\beta}$ and
$\omega^{(-)ab}$ becomes $\omega^{\adot\bdot}$.

An equivalent form of  the Einstein-Hilbert
action~(\ref{EH1order}) is given using $R^{(+)}$ instead of $R$
by
\begin{equation} \frac{1}{2\kappa^2} \int    e^a \wedge e^b
\wedge R^{(+)}_{ab} (\omega ) . \label{EH1order+}
\end{equation}
This gives the action~(\ref{EH1order}) plus the topological term
\begin{equation}
\frac{1}{2\kappa^2} \int   \,  e^a \wedge e^b \wedge R_{ab}
(\omega ) . \label{Hir1}
\end{equation}
Using~(\ref{Texpl}) and~(\ref{R2form}) this can be written as
\begin{equation}
\frac{1}{2\kappa^2} \int   \,  d(T^a  \wedge e_a) , \label{Hir2}
\end{equation}
which vanishes in the second order formalism in which one sets
$T^a=0$, and in the first order formalism is a total derivative
that  does not contribute to the field equations or Feynman
diagrams. As $R^{(+)}$ depends only on $\omega ^{(+)}$, the
action~(\ref{EH1order+}) is independent of $\omega ^{(-)}$ and
depends only on the vierbein and the self-dual spin-connection.
Moreover, the first order action is polynomial in these variables.

The form~(\ref{EH1order+}) of the action
has been used as a covariant  basis for the
reformulation of general relativity in terms of Ashtekar
variables,
and can
 be rewritten in two-component spinor notation as~\cite{Capo1,Capo2}
 \begin{equation}
\int  \Sigma^{\a\b} \wedge R_{\a\b} -\frac{1}{2}\psi_{\a\b\c\d}
\Sigma^{\a\b} \wedge \Sigma^{\c \d}   \label{Cap}
\end{equation}
where   the curvature
2-form $R_{\a\b}$ of $\omega^{\a\b}$ is given in eq.~(\ref{Rsd})
and $\Sigma$
is a a self-dual 2-form acting as a Lagrange multiplier.
The totally symmetric Lagrange multiplier field $\psi_{\a\b\c\d}$ imposes the constraint
 \be
 \Sigma^{(\a\b}\wedge \Sigma^{\c\d)}  =  0
\ee
which implies that
\begin{equation}
\Sigma^{\a\b} = e^{\a}{}_{\adot}\wedge e^{\b\adot} \label{sole}
\end{equation}
for some tetrad $e^{\a\adot}$.  Solving for $\Sigma^{\a\b}$ as in~(\ref{sole}) and
substituting in~(\ref{Cap})  yields~(\ref{EH1order+}).

It is remarkable that one only needs the self-dual part of the
spin-connection in order to  formulate gravity. The torsion
constructed from $e, \omega ^{(+)}$ is (setting  $\omega ^{(-)}=0$
in~(\ref{Texpl}))
\begin{equation}\label{Texpl+}
  \tilde T^a =   d e^a +
  \omega ^{(+)a}{}_b \wedge
  e^b .
   \end{equation}
If one imposes the constraint $\tilde T^a = 0$, one obtains \be
\omega^{(+)ab}=\Omega^{(+)ab}(e) \label{omOm+}\ee and \be
\Omega^{(-)ab}(e)=0 .\label{0Om-}\ee Now~(\ref{0Om-}) implies in
turn that \be R^{(-)ab}(e)=0 ,\ee where $R^{(-)ab}(e)$ is the
anti-self-dual part of the curvature of the connection
$\Omega(e)$. Then the Riemann curvature constructed from the
vierbein is self-dual and hence Ricci-flat, so that the torsion
constraint $\tilde T^a=0$ imposes the field equations of self-dual
gravity as well as solving for the spin-connection in terms of the
vierbein ~\cite{Siegel2}.

Siegel~\cite{Siegel2} gave a remarkable asymmetric action for gravity
that is analogous to the asymmetric gauge theory
action~(\ref{SiegelYM1}) by introducing a Lagrange multiplier
field to impose the constraint $\tilde T=0$. In the second order
formalism, $\omega^{(+)ab}$ is given in terms of $e$
by~(\ref{omOm+}) and the remaining part of $\tilde T^a=0$ is
imposed by a Lagrange multiplier $\s_\mu {} ^{(-)ab}$ which is
anti-self-dual, or in spinor notation $\s_\mu  {} ^{\adot\bdot}$.
This has the same index structure as the missing anti-self-dual
spin-connection. Siegel's action can be written as \be \int \, \s
^{\adot\bdot} \wedge \tilde T^\a {}_{\adot} \wedge e_{\a\bdot} .
\label{siegold} \ee Varying $\s$ imposes the self-dual gravity
equation~(\ref{0Om-}) so that $e$ represents a graviton of
helicity $-2$. Varying $e$ gives \be  d \sigma^{\adot \bdot}
\wedge e_{\a\bdot}= 0 ,\ee so that the Ricci tensor constructed from the
linearised curvature $d \sigma$ for an
anti-self-dual connection $\s$ vanishes, and the Lagrange
multiplier field represents a graviton of helicity $+2$. This
action then represents particles of helicity $\pm 2$, as in
Einstein's theory, but the interactions are different for the two
helicities, and in particular the theory is linear in $\s$.

There
is also a first-order form of this theory, in which
$\omega^{(+)ab}$ is an independent field and a Lagrange multiplier
is introduced to impose the full constraint $\tilde
T=0$~\cite{Siegel2}, which in turn implies the field equations~\cite{Siegel2}.  Siegel also generalised~(\ref{siegold}) to give
an asymmetric form of $N=8$ supergravity, with Lagrange multipliers
imposing torsion constraints of the supergravity
theory~\cite{Siegel2}.

Siegel's asymmetric  theory of gravity can be put in a different form
that arises as a  weak-coupling limit of the Einstein theory, and
gives a  chiral perturbation theory of gravity ~\cite{BerkSieg,fields} similar to that arising
from the Yang-Mills action~(\ref{SiegelYM2}). The gravity
action~(\ref{EH1order+}) depends on the vierbein and
$\omega^{(+)}$ only, and
  $\omega^{(-)}$ decouples completely: we can write it in the form
\begin{equation}
 \frac{1}{2} \int  e^a \wedge e^b \wedge
 \left( d\omega_{ab} +\kappa^2 \omega_{ac} \wedge \omega ^{c}{}_b \right)
  \label{EH1order9} ,
\end{equation}
where from now on we omit the superscript ${(+)}$ so that $\omega \equiv \omega^{(+)}$ and we have rescaled the
connection by the gravitational coupling $\kappa^2$.
Varying~(\ref{EH1order9}) independently with respect to
$\omega^{a}{}_b$
  and   $e_\mu ^a$, we obtain~(\ref{omOm+}) giving the
 connection in terms of the vierbein,
and the Einstein equation
\begin{equation}
e^a\wedge \left( d\omega_{ab} +\kappa^2 \omega_{ac} \wedge \omega
^{c}{}_b \right)=0 . \label{eR0}
\end{equation}

Now  taking the limit $\kappa \rightarrow 0$ in~(\ref{EH1order9})
yields a  weak-coupling limit of gravity with action
\begin{equation}
  \frac{1}{2} \int   e^a \wedge e^b \wedge d \omega_{ab}
  \label{EH1order10} ,
\end{equation}
which can be rewritten using eq.~(\ref{usom}) as
\begin{equation}
- \int
  e^a \wedge e^b \wedge
\omega_{ac} \wedge \Omega^{c}{}_b=
 - \int
  e^a \wedge e^b \wedge
\omega_{ac} \wedge \Omega^{(+)c}{}_b
  \label{acts+}
\end{equation}
where $\Omega^{(+)} =\Omega^{(+)} (e)$ is the self-dual part of
the connection~(\ref{usom}). This is an action for two independent
fields, the vierbein $e_{\mu}{}^a$ and the self-dual connection
$\omega ^{ac}$; the latter now plays the role of a Lagrange
multiplier field. Note that the self-duality of $\omega ^{ac}$
implies that only the self-dual part $\Omega ^{(+)}$ of $\Omega
(e)$ occurs in the action.
 The field
equation from varying the Lagrange multiplier field
$\omega ^{ac}$ sets the self-dual part of $\Omega(e)$
to zero,
\begin{equation}
\Omega^{(+)a}{}_b (e) = 0 . \label{Omplus0}
\end{equation}
This implies that the self-dual part of the curvature constructed
from the Levi-Civita connection $\Omega (e)$ vanishes
\begin{equation}
R_{\mu\nu}{}^{ab} (\Omega^{(+)}) = 0 ,
\end{equation}
so that the vierbein gives a metric with anti-self-dual Riemann
curvature. The field equation for the vierbein gives
\begin{equation}
e_b \wedge d\omega ^{ab} = 0 . \label{eqforsig}
\end{equation}
Comparing with~(\ref{eR0}), this can be seen to be a version of
the Einstein equation linearised around the anti-self-dual
background spacetime described by the tetrad $e^a$, where the
linearised graviton field is the self-dual connection
$\omega ^{ac}$.

The fact that
$\omega ^{ab}$ and $\Omega^{(- )} (e)$
  are respectively self-dual and anti
self-dual   means that they describe particles of opposite
helicity: $e$ describes a particle of helicity $+2$ and $\omega $
describes a particle of helicity $-2$. The linearized spectrum is
the same for~(\ref{EH1order10}) and~(\ref{EH1order}), but the
interactions differ as~(\ref{EH1order10}) has no $--+$ vertex. The
asymmetric theory~(\ref{EH1order10}) is equivalent to the Siegel
theory (but with opposite conventions to those used earlier;  in~(\ref{siegold}) $e$
represents a negative helicity graviton, while
in~(\ref{EH1order10}) it is a positive helicity one).

Now the form of the action~(\ref{EH1order9}) has the weak-coupling
limit~(\ref{EH1order10}), and one can consider perturbation theory
in $\kappa^2 $ about this weak-coupling limit, in complete analogy
with that of the gauge theory~(\ref{SiegelYM2}). As in that case,
it is useful to attribute to the independent fields $e^{a}$
and $\omega ^{ab}$ the weights $w[e]=0 $ and $w[\omega ]=-1$
under a $U(1) $ transformation, corresponding to the
  \lq anomalous' $U(1)$ R-symmetry $S$
in the $N=8$ supersymmetric extension of
action~(\ref{siegold})~\cite{Siegel2}, with $S=8w$. The
 chiral action~(\ref{EH1order+}) has
weight $w=-1$ and the second term in~(\ref{EH1order9}) has weight
$w=-2$.  It is tempting to conjecture that the $S=-16$ interaction
term is related to nonperturbative contributions in a new
twistor-string theory for Siegel's truncation of $N=8$
supergravity. A similar conjecture was made in~\cite{Nair2} based
on an analysis
 of maximally helicity violating scattering amplitudes for gravitons.

Consider the perturbation theory in $\kappa^2$ for the theory
defined by action~(\ref{EH1order9}). If one attributes weights
$w=1$ to $\kappa^2$ and $w=-1$ to the Planck constant $\hbar$,
then the action rescaled by $1/ \hbar$ has weight $w=0$. The
generating functional of
  scattering matrix elements at $l$-loops must be a sum of terms of the form
  \be
\hbar^{l-1} \tilde{f} (e)  \kappa^{2d} \omega^q \ee for some
function $\tilde{f}$, and for this to have total weight zero it is
again necessary that the relation~(\ref{Wittform}) holds for each
term in the effective action.  The power $q$ of $\omega$ is the
number of negative helicity gravitons in an $l$-loop scattering
process in the theory defined by~(\ref{EH1order9}). If the theory
has a twistor  string origin similar to that of~\cite{Witten2003},
then the power $d$ of $\kappa^2$ might arise as  an instanton
number, and the scattering would have support on   curves in
twistor space characterised by the integer $d$.

This formulation of gravity extends to one of $N=8$ supergravity
in which the Einstein term is written in the
form~(\ref{EH1order9}) and the vector field kinetic terms take the
form~(\ref{SiegelYM2}). In the weak coupling limit, it gives
Siegel's chiral $N=8$ supergravity~\cite{Siegel2}.

We have seen that the formulation of gauge theory with
action~(\ref{SiegelYM2}) and that of gravity with
action~(\ref{EH1order9}) have many similarities: both have a
non-trivial asymmetric weak-coupling limit and both have a
  perturbation theory that leads to the relation~(\ref{Wittform}). For the gauge theory,
 the action~(\ref{SiegelYM2}) arises from a twistor
string theory with the first term arising from the perturbative theory and the second term from instanton corrections.
The constraint~(\ref{Wittform}) then implies that
amplitudes are supported on holomorphic curves of degree $d$ in
twistor space. It is natural to conjecture that the gravity
action~(\ref{EH1order9}), or its $N=8$ supergravity generalisation,  should also have an elegant twistor
theory origin, and that the formula~(\ref{Wittform}) has a similar
twistor space interpretation. We will discuss the twistor formulation of this theory elsewhere.

\section*{Acknowledgements}

We would like to thank  Lionel Mason and Prameswaran Nair  for
helpful discussions. We also thank the Vrije Universiteit Brussel
and the Institute for Mathematical Sciences at Imperial College
London for hospitality and support. The work of M.\ A.\ was
supported by a PPARC Postdoctoral Research Fellowship with grant
reference PPA/P/S/2000/00402, by the \lq FWO-Vlaandere' through
project G.0034.02, by the Belgian Federal Science Policy Office
through the Interuniversity Attraction Pole P5/27, by the European
Commission FP6 RTN programme MRTN-CT-2004-005104, by the European
Commission through its 6th Framework Programme \lq Structuring the
European Research Area' and by contract Nr. RITA-CT-2004-505493
for the provision of Transnational Access implemented as Specific
Support Action.

\end{document}